# Continuous Wave Photonic Crystal Laser in Ultraviolet Range

Y.V. Radeonychev and I.V. Koryukin

*Institute of Applied Physics, Russian Academy of Science,
46 Ulyanov Str., 603950 Nizhny Novgorod, Russia*

**Abstract**

Great potential of photonic band-gap (PBG) structures for the development of continuous wave ultraviolet all-solid-state laser, including UV femtosecond pulse source, as well as for increase of output laser power at short-wave optical range is shown. The basic idea is a decrease of the laser threshold via suppression of the spontaneous emission in the active medium by means of embedding it into 2D or 3D PBG structure having the midgap frequency close to the frequency of the laser transition. Different schemes of 2D and 3D photonic crystal laser for ultraviolet range are proposed.

1. Widely tunable ultraviolet (UV) coherent radiation is important for applications in various fields including physics, chemistry, biology and engineering. Production of broadband continuous wave (cw) UV laser fields is especially important since it would allow generation of ultrashort UV pulses that is one of the needs of ultrafast diagnostics and various potential applications. At present, the available methods for getting UV coherent radiation are not able to answer these needs. No sources of cw broadband laser radiation in UV range do exist. Thereupon development of solid-state broadband and widely tunable cw UV lasers is one of the topical goals of laser science.

Radiative spontaneous relaxation of a quantum system is a fundamental obstacle for UV lasing. Quantum electrodynamics provides the most adequate treatment of the spontaneous emission as a result of interaction of a quantum system (atom, ion, molecule, quantum dot) with electromagnetic field in the vacuum state. Spontaneous emission rate of a quantum system (the Einstein coefficient $A$) in free space is commonly expressed as [1]:

$$A = \frac{4\omega^3 d^2}{3\hbar c^3}, \qquad (1)$$

where $\omega$ is the angular frequency of quantum transition, $d$ is modulus of a matrix element of a dipole moment, $\hbar$ is Plank constant and $c$ is the light velocity in vacuum. Cubic dependence of the spontaneous emission rate on the transition frequency implies fast radiative relaxation at high frequencies. Intensive spontaneous emission determines short lifetime of the upper laser energy level. This in turn embarrasses to create the required value of population inversion for laser gain preventing lasing in the UV range. As a result, most of the solid-state lasers oscillate in infrared region and long-wave part of the optical domain. The high-frequency region is not occupied in spite of the fact that the transparency range for the most of laser matrices covers the UV domain, expanding beyond 200 nm, as well as active doping ions, having necessary transitions, exist.

It is well known that the spontaneous emission rate (1) is the result of integration over all possible electromagnetic modes of the field reservoir, interacting with the quantum system. In more general form it can be written as

$$A = \frac{2\pi}{\hbar^2} \int \left|V(\vec{k})\right|^2 \delta(\omega - \omega(\vec{k})) D(\vec{k}) d\vec{k}, \qquad (2)$$

where $\vec{k}$ is the wave vector of electromagnetic mode, $V(\vec{k})$ is the matrix element of the interaction Hamiltonian of quantum system with electromagnetic mode, $\delta$ is the Kronecker symbol, $\omega(\vec{k})$ is the frequency of the electromagnetic mode, $D(\vec{k})$ is the density of electromagnetic modes.

Purcell was the first who took note of the dependence of the spontaneous emission rate (2) on the density of electromagnetic modes of an environment [2]. He predicted that the spontaneous emission of the excited atom can be drastically altered when it is placed into a high-Q cavity with dimensions of the order of the emitted wavelength (the Purcell effect). This is because the mode density in such a cavity differs from the mode density in free space. Later Bykov predicted possibility of reduction of the spontaneous emission rate inside a one-dimensional (1D) structure having periodic spatial variation of the refractive index [3]. Yablonovitch showed that dielectric structures with three-dimensional (3D) spatial periodic variation of the refractive index, called photonic crystals (PCs), can exhibit ranges of frequencies, for which electromagnetic wave propagation is forbidden (photonic band gaps). Such structures can demonstrate complete suppression of spontaneous emission within the photonic band gap (PBG) due to zero value of the mode density [4]. In the vicinity of the PBG edge amplification of the spontaneous emission can also occur due to higher value of the mode density [5]. Nowadays two dimensional (2D) PBG structures are much easier to fabricate than the 3D structures and the most of publications are devoted to them. In the 2D PCs mode density in (2) strongly depends on the direction of the wave vector. The PBG can exist only for radiation that propagates in the plane of PC. Therefore 2D PCs can have only pseudogap [6-11] and can provide only partial suppression of spontaneous emission. For quantitative evaluations, we introduce spontaneous emission suppression (SES) factor,

$$F = \frac{A_{PC}}{A_{FS}}, \qquad (3)$$

as a ratio of spontaneous emission rate in a photonic crystal, $A_{PC}$, to the spontaneous emission rate in free space, $A_{FS}$. Then the 2D PCs can have $0 < F < 1$ and only the 3D PCs are able to provide $F=0$.

PBG structures are mostly investigated in semiconductor media [8-11]. Semiconductors have high value of the refractive index $n$ necessary for PBG formation, fabrication technology of semiconductor structures is highly developed and continuously improved. In addition to study of nonlinear properties, semiconductor PBG structures are investigated mainly for getting microcavities and increase of the spontaneous emission for microlasers and single-photon sources [8,9,12,13]. Except for unique cases [14] semiconductor PCs normally operate in the long-wave domain. Interband absorption of UV photons, inevitable in semiconductors, hampers use of semiconductor PCs in the UV range.

In this paper, we consider the dielectric PBG structures for the development of UV laser as well as for increase of output laser power at the short-wave optical range. The basic idea is a decrease of the laser threshold via suppression of the spontaneous emission of the active particles (an increase of the lifetime of the upper laser level) by means of imbedding the active medium into PBG structure.

2. The 3D PCs provide the best conditions for realization of cw ultraviolet laser (Fig.1). A proper sample can be fabricated either from some dielectric material (Fig.1a) or directly from the active medium (Fig.1b). It should have properly constructed periodic array of holes or another kind of the refraction index spatial periodicity such that a PBG would exist having the midgap frequency $\omega_m$ localized near the frequency of the laser transition. For widely tunable lasing, PBG should have enough width $\Delta\omega$ in order to cover broad emission line of the laser transition. For the case of $Ce^{3+}$ discussed below one needs $\Delta\omega/\omega_m \sim 10\%$. This condition should be met by optimal combination of geometry and period of PC structure, filling ratio, and dielectric contrast. The laser cavity in PBG material can be formed by means of a linear waveguide defect ended by the mirrors. In the case where PC is made from the active material waveguide defect can be created by missing several rows of holes (Fig.1b) that leads to formation of donor modes [14, 15]. It could be more expedient to fabricate PC structure from a suitable dielectric material different from the active medium. In this case, a laser crystal rod surrounded by a dielectric PC structure makes up the needed device. The waveguide defect in such a hybrid construction has acceptor modes (Fig.1a).

In the presence of the waveguide defect total SES factor $F_{tot}$ is sum of SES factor in PC, $F_{PC}$, and SES factor in the waveguide, $F_{WG}$,

$$F_{tot} = F_{FS} + F_{WG} = \frac{A_{PC}}{A_{FS}} + \frac{A_{WG}}{A_{FS}} \qquad (4)$$

where $A_{WG}$ is spontaneous emission rate into the waveguide. The waveguide SES factor $F_{WG}$ can be roughly estimated as a ratio of a solid angle $\Omega$, in which spontaneous emission takes place, to the solid angle $4\pi$ of whole space, $F_{WG}=2\Omega/(4\pi)$ (Fig.1). It is valid in the case of isotropic distribution of directions of the dipole moment vectors of active particles. For minimal defect diameter $d\sim\lambda$ one has,

$$F_{WG} \sim \frac{\lambda^2}{2L^2} \qquad (5)$$

where $L$ is length of a sample. For $\lambda=300$ nm and $L=0.3$ mm, one has $F_{WG}\sim 5*10^{-7}$. This means that even for miniature samples waveguide defect gives negligible contribution to $F_{tot}$.

The cavity mirrors could be made as multilayer dielectric mirrors selectively reflecting at the laser frequency and mounted at the ends of the sample (Fig.1a). They also could be a part of PC structure (Fig.1b). Optical pump of the active medium can be realized from any direction (including the mirrors) since the pump frequency is normally outside the PBG. In the case where PBG structure is made from the laser material (Fig.1b) end pump is preferable to avoid parasitic resonant absorption by the active particles outside the waveguide defect.

It should be highlighted that *there is no need to have perfect PBG*. As shown below, a pseudogap providing suppression of spontaneous emission as much as several times can be enough for UV lasing. The most of PC geometries with PBGs have pseudogaps. This means that some directions within PBG remain more or less allowed for light propagation. One of such directions in properly designed PC geometry can serve for creation a laser cavity instead of a waveguide defect. Induced emission will propagate in this direction to the mirrors.

3. In the case of 2D PC, pseudogap is only possible. Only a part of spontaneous radiation emitted in the plane of PC slab can be suppressed. Nevertheless, the 2D PCs are promising for UV lasing since there is no need to have very small value of the SES factor. In the experiments with InAs quantum dots imbedded in the GaAs 2D PC fivefold increase of the exited state lifetime was demonstrated [9]. Theoretical calculation of a similar PC sample with finite dimensions by the finite difference time domain (FDTD) method showed possibility of the sevenfold suppression of the spontaneous emission [16] in the case of optimal orientation and position of the emitting dipole in a PC cell. Maximal suppression of spontaneous emission occurs for dipoles oriented in the plane of PC (polarization of the TE modes) and situated inside the material (not inside the holes) [17, 18]. Both of these points are favorable for the PC-based laser (Fig.2).

Similar to the case of the 3D PCs either a waveguide defect or a specific direction allowed for light propagation in the 2D geometry can be used for the laser cavity. A waveguide defect can be formed by missing several rows of holes in a slab if the slab is made from a laser material (Fig.2a). As was shown in [19], the acceptor TE modes of such a defect have the frequencies that fall just inside the photonic pseudogap. Alternatively, a hybrid construction, similar to the 3D case (Fig.1a), fabricated from an active medium surrounded by a proper dielectric PC structure can form the needed device. If the refractive index of a slab exceeds the refractive index of a substrate, a common dielectric waveguide exists in the third direction. Properly constructed ends of the slab (multilayer coating or a part of PC structure discussed above) can serve as laser mirrors.

Another laser scheme based on the 2D PC structure made from the active material can be proposed (Fig.2b). In this case, stimulated emission goes on in vertical direction in the absence of a waveguide defect. Mirrors coated on the top and bottom planes of a slab form the laser cavity.

The discussed above 3D and 2D variants of PC-based laser can serve as a basis for development of miniature monoblock all-solid-state UV chip-lasers with cw broadband or widely tunable output. Besides, they could be used for lasing at novel frequencies as well as for increase of output power of lasers in high-frequency visible range.

4. It is known that the perfect PBG can exist on condition that the refractive index contrast exceeds some threshold value. In the case of 3D PC structure and the most common pair dielectric–air the perfect PBG requires a material with refraction index $n>2$ [15, 20]. Otherwise, only a pseudogap can exist. A number of laser crystals matrices, e.g. $Gd_2O_3$, $La_2O_2S$, $LiNbO_3$, and other niobates have $n>2$ [21]. However because one never needs the perfect PBG, but a pseudogap only, every laser crystal can be considered as a potential material for PC laser discussed above. It should also be noted that in the 2D case the requirement for refraction index contrast is much softer than in the 3D case [20]. While fabrication of 3D PBG structures of sub-micron size is technologically challenging, the state-of-the-art technologies (e.g. focused ion beam) do allow producing different 2D nanometer-sized structures in dielectric crystals [22].

The most promising candidate for UV cw lasing is triple-ionized cerium doped into some solid-state host [23]. At present trivalent cerium-doped fluorides are the only solid-state materials that provide direct lasing in UV range. Fortunately, cerium-doped lasers demonstrate high efficiency and wide tunability (~10%) in the 280-330 nm range, as well as effectiveness of frequency doubled/quadrupled visible/IR solid-state pump, thus providing an all-solid-state route for femtosecond UV pulse generation. The two most-used lasers are based on $Ce^{3+}:LiLuF_4$ and $Ce^{3+}:LiCaAlF_6$ (Ce:LiCAF). The last one demonstrates the highest operation efficiency.

Unfortunately, all the cerium-doped lasers are restricted by pulsed-mode operation because of insufficient cw pump power. For the case of Ce:LiCAF the estimated cw threshold pump power about five times exceeds the maximum cw power available by commercial devices [24]. The major factor that determines high threshold pump power for solid-state UV lasing (aside from fast exited state radiative relaxation) is the exited state absorption (ESA) to the conduction band of the host crystal. If the ESA cross section $\sigma_{ESA}$ exceeds the stimulated emission cross section $\sigma_{em}$, no lasing is possible. Theoretical estimation of the overall threshold pump power $P_{th}^{cw}$ that takes into account various experimental factors [24] shows that

$$P_{th}^{cw} \sim \frac{A}{\sigma_{em} - \sigma_{ESA}}, \quad (6)$$

where $A$ is spontaneous emission rate from the upper laser level. It follows from (6) that in the case of Ce:LiCAF fivefold suppression of spontaneous emission rate due to PC structure can provide cw lasing with commercially available pump sources.

Since the laser output $P_{out}$ is proportional to the pump power $P_{pump}$,

$$P_{out} = \eta \left( P_{pump} - P_{th}^{cw} \right) \quad (7)$$

(where $\eta$ is the slope efficiency) then suppression of the spontaneous emission by PC structure leads to increase of output laser power.

There is a common rule of laser physics stating that high value of population inversion $N$ can be reached only at quantum transitions that have small oscillator strength (contain metastable states). However, this implies small value of the stimulated emission cross section $\sigma_{em}$ and can result in small laser gain $G$, because $G = \sigma_{em} N$. These two factors give rise to the well-known dilemma of laser gain: large $N$ at the price of small $\sigma_{em}$ or large $\sigma_{em}$ at the price of small $N$. Use of PC structures can revise the above rule and soften the dilemma since suppression of the spontaneous emission leads to increase of both $N$ and $\sigma_{em}$.

We have shown that PCs have great potential in short-wave laser physics and are a promising tool for the development of UV femtosecond pulse source.

This research was supported by RFBR grant 06-02-16632, ISTC grant A-1095, CRDF grant RUP2-2844-NN-06, and the Russian President Program for Support of Leading Scientific Schools (grant 7738.2006.2).

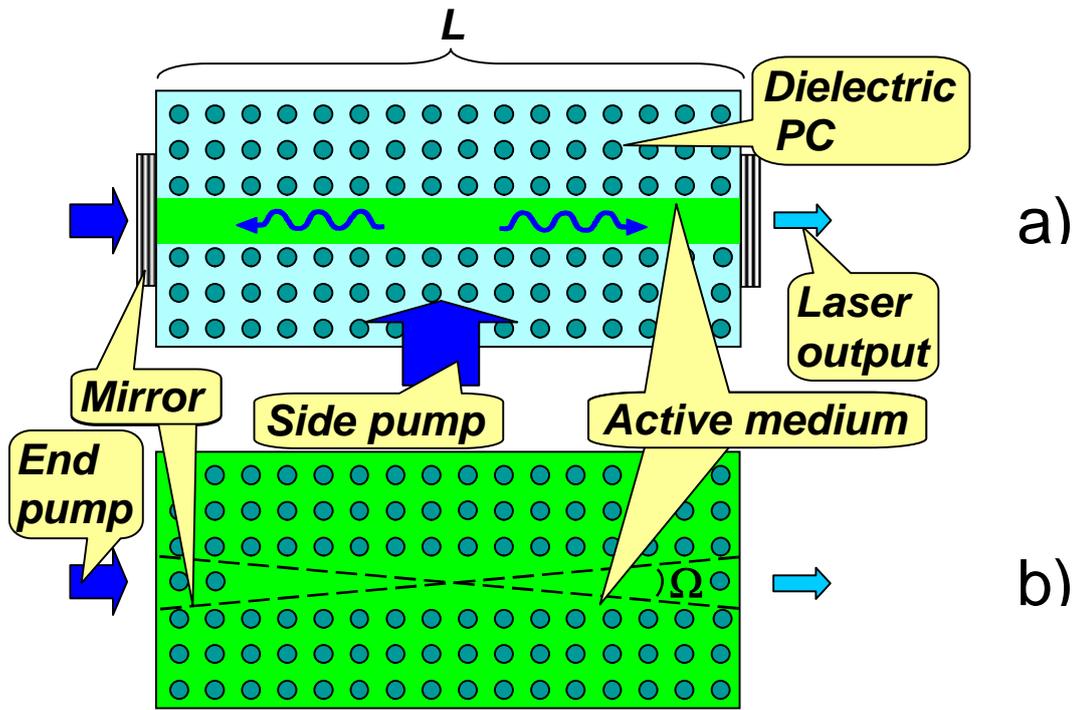

Fig. 1. (color online) Schemes of 3D PC-based laser, in section view: a) a hybrid construction; b) PC is made from the active medium.

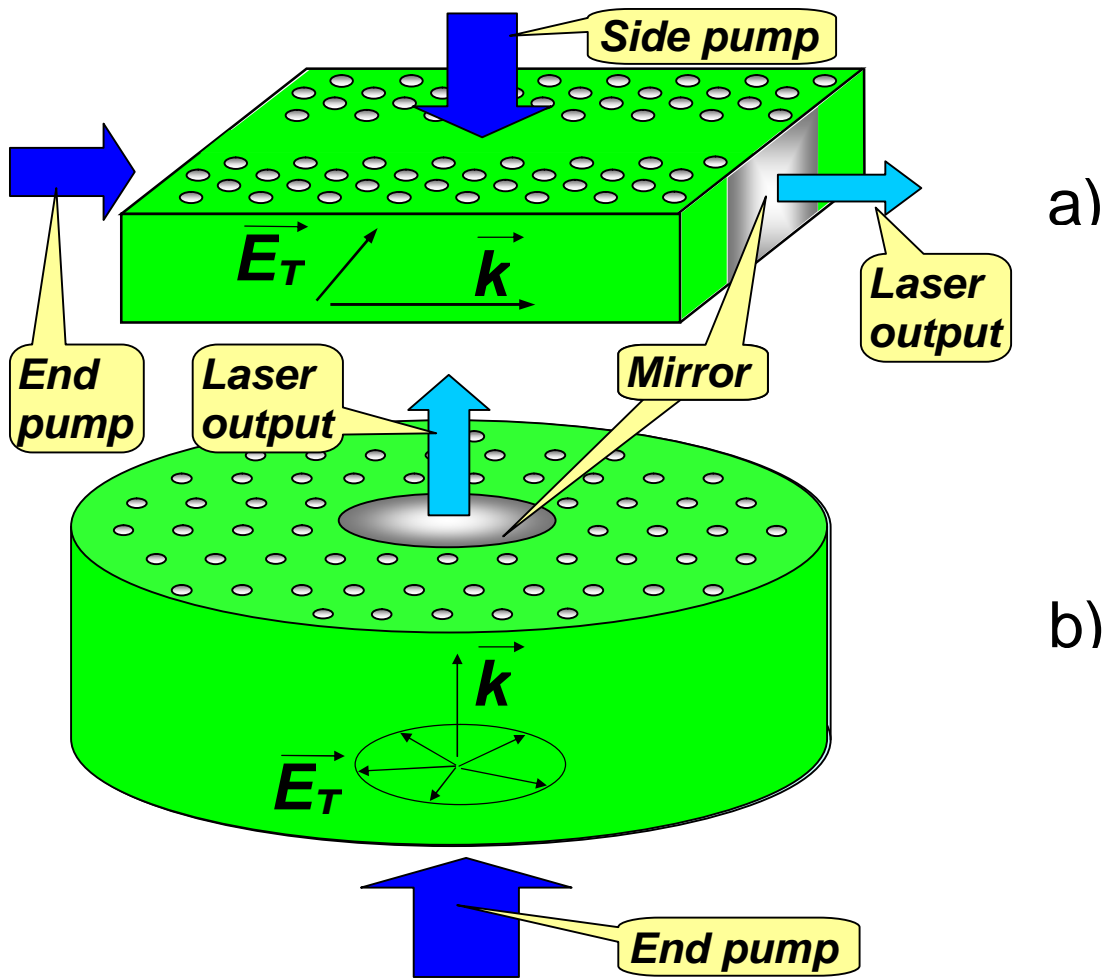

Fig. 2. (color online) Schemes of 2D PC-based laser made from the active medium: a) with waveguide defect; b) with vertical output.